\documentstyle[letter]{ptptex}
\title{Comments on the Gravitational Lensing Magnification}
\author{Takashi \sc{Hamana} }
\inst{Astronomical Institute, Tohoku University, Sendai 980-8578, Japan}
\recdate{\today}
\abst{We rederive a relation between gravitational lensing
magnification relative to the standard Friedmann distance and one
relative to the Dyer-Roeder distance by investigating the null geodesic
deviation equation. We show that the relation comes from a natural
consequence of the definition of the lensing magnification matrices
and is not based on the averaging of the magnifications, which has
conventionally been used to derive it. We therefore conclude that the
relation is true for each individual ray bundle.}
\begin{document}
\maketitle
\section{Introduction}
Since the discovery of the first multiply imaged quasars \cite{wa79}
and the first observations of gravitational arcs \cite{ly86,so87} and
arclets, \cite{fo88} gravitational lensing has rapidly become one of the
most promising tools for cosmology. In studies of gravitational
lensing, the lensing magnification plays an important role, such as
the magnification bias in the statistical study of multiply imaged
quasars \cite{fu91,ha97} and the number count of distant galaxies in
the field of cluster of galaxies.~\cite{fo97}

Conventionally, there are two definitions of the lensing magnification,
namely, the magnification relative to the smooth Friedmann distance and
that relative to the Dyer-Roeder distance. The relation between these two
was derived from the averaging of magnifications over an ensemble of
sources based on the argument of flux conservation (see, e.g., Section
4 of Ref.~\citen{sc92}). Therefore, it is not clear
whether the relation is true for each individual ray bundle.

The main purpose of this paper is to show that this relation
results from a natural consequence of the definition of the lensing
magnification matrices. Accordingly, we find that the relation is true
for each individual ray bundle. 

Throughout this paper, we use units for which $c=H_0=1$ and the
scale factor $a$ is normalized to be unity at the present epoch
($a_0=1$). The density parameter $\Omega_0$ and normalized
cosmological constant $\lambda_0$ are defined in the usual manner.

\section{Basic equations}
The propagation of a bundle of light rays in an inhomogeneous universe
was investigated in Refs.~\citen{sa93} and \citen{se94} in detail. 
In this section,
we simply describe only the aspects which are directly relevant to
this paper. 

\subsection{Universe model}
It is well known that the metric of a realistic model of our universe is
well approximated by the cosmological Newtonian metric of the form
\cite{fu88,fu89}
\begin{eqnarray}
\label{metric}
d\hat{s}^2 &=& a^2(\eta)\left[ -(1+2\phi) d\eta^2 + (1-2\phi)\gamma_{ij}
dx^i dx^j \right],\\
\label{gamma}
\gamma_{ij} dx^i dx^j &=& d\chi^2+f^2(\chi) \left(d\theta^2 +\sin^2\theta
d\varphi^2 \right),\\
\label{f}
f(\chi) &=& \left\{
\begin{array}{ll}
{K}^{-1/2}\sin(\sqrt{K} \chi) & \mbox{for {} } K>0\\
\chi & \mbox{for {} } K=0 \\
(-K)^{-1/2} \sinh(\sqrt{-K} \chi) & \mbox{for {} } K<0,
\end{array}
\right.
\end{eqnarray}
where $\eta$ is a conformal time and $\gamma_{ij}$ is the metric on
the constant curvature 3-space with curvature
$K=\Omega_0+\lambda_0-1$. The scale factor $a$ and Newtonian potential
$\phi$ are determined by the following equations to the lowest order:
\begin{eqnarray}
\label{scale}
\left( {{a'} \over {a}} \right)^2 &=& {{\Omega_0} \over a} - K +
\lambda_0 a^2,\\
\label{pois}
 \Delta^{(3)} \phi &=& 4 \pi G ( \rho-\rho_b) a^2 = {3 \over 2} {{\Omega_0}
\over a} \delta.
\end{eqnarray}
Here ${}'\equiv d/d\eta$, $\Delta^{(3)}$ is the Laplacian operator 
in the spatial section, $\rho_b$ is a mean matter density, and $\delta$
is the density contrast defined by $\delta \equiv
(\rho-\rho_b)/\rho_b$. We write the above metric as $\hat{g}_{\mu
\nu} = a^2 g_{\mu \nu}$. Since the light cone structure is invariant
under the conformal transformation of the metric, in the following we
work in the conformally related $g$ world.

\subsection{Propagation of a bundle of light rays}
Let us consider an infinitesimal bundle of light rays
intersecting at the observer. We denote a connecting vector
which connects the fiducial light ray $\gamma$ to one of its neighbors 
as $\xi^\mu$.
All gravitational focusing and shearing effects on the infinitesimal
light ray bundle are described by the geodesic deviation equation,
\begin{equation}
\label{geodev}
{{d^2 \xi^{\mu}} \over {d \lambda^2}} = -{R^{\mu}}_{\alpha \nu \beta}
\xi^{\nu} k^{\alpha} k^{\beta},
\end{equation}
where $k^{\alpha}=dx^{\alpha}/d\lambda$, and $\lambda$ is the affine
parameter along the fiducial light ray $\gamma$. We introduce a dyad
basis $e_{\mu}{}^A$ ($A$, $B$, $C,...=1,2$) in the two-dimensional screen
orthogonal to $k^{\mu}$ and parallel-propagated along $\gamma$.
The screen components of the connection vector are given by
\begin{equation}
\label{Y}
Y^A = e_{\mu}{}^A \xi^{\mu}.
\end{equation} 
From the geodesic deviation equation (\ref{geodev}), one can
immediately find that $Y^A$ satisfies the Jacobi differential equation
\begin{equation}
\label{jaco}
{{d^2 Y_A}\over {d\lambda^2}} = {\cal{T}}_{A B} Y^B,
\end{equation}
here ${\cal{T}}_{AB}$ is the so-called optically tidal
matrix.~\cite{se94} From the metric (\ref{metric}), up to first order
in $\phi$, this matrix is given by 
\begin{equation}
\label{tid}
\mbox{{\boldmath${\cal{T}}$}} = 
-K \mbox{{\boldmath${\cal{I}}$}} - \left(
\begin{array}{ll}
{\cal{R}} + \mbox{Re}\left[{\cal{F}}\right] &
\mbox{Im}\left[{\cal{F}}\right] \\
\mbox{Im}\left[{\cal{F}}\right] &
{\cal{R}} - \mbox{Re}\left[{\cal{F}}\right]
\end{array}
\right),
\end{equation}
where {\boldmath${\cal{I}}$} is a $2 \times 2$ identity matrix, and 
\begin{eqnarray}
\label{Ricci}
{\cal{R}} &=& \Delta^{(3)} \phi,\\
\label{Wely}
{\cal{F}} &=& \phi_{,11}-\phi_{,22} + 2 i \phi_{,12}
\end{eqnarray}
represent the Ricci and Weyl focusing induced by the density
inhomogeneities, respectively. By the linearity of (\ref{jaco}), 
the solution of $Y^A$ is written in terms of its initial value
$dY^A/d\lambda |_{\lambda=0} = \vartheta^A$ and the $\lambda$-dependent
linear transformation matrix ${\cal{D}}_{AB}$ as 
\begin{equation}
\label{D}
Y^A(\lambda) = {\cal{D}}^A{}_B(\lambda) \vartheta^B.
\end{equation}
Substituting the last equation into the Jacobi differential equation
(\ref{jaco}), we obtain 
\begin{equation}
\label{evoD}
{{d^2{\cal{D}}_{AB}} \over {d \lambda^2}} = {\cal{T}}_{AC} {\cal{D}}_{CB}.
\end{equation}
This is our principal equation.

\section{Lensing magnifications}
Now, we derive an evolution equation of the lensing magnification
matrix relative to the smooth Friedmann distance from (\ref{evoD}). 
First, we write (\ref{tid}) as
{\boldmath${\cal{T}}$}$ = ${\boldmath${\cal{T}}$}${}^{(0)} +
${\boldmath$\delta{\cal{T}}$}, with {\boldmath${\cal{T}}$}${}^{(0)} =
-K${\boldmath${\cal{I}}$}, and {\boldmath$\delta {\cal{T}}$} is the
second term in (\ref{tid}). In the homogeneous case,
{\boldmath$\delta{\cal{T}}$} is vanishing, and the solution of
{\boldmath${\cal{D}}$} is ${\cal{D}}_{AB}(\lambda) = D_f(\lambda)
\delta_{AB} = f(\lambda) \delta_{AB}$, where $D_f$ is, of course, the
standard angular diameter distance in the background Friedmann
universe. 

It is natural to define the lensing magnification matrix relative to the
corresponding Friedmann universe as
\begin{equation}
\label{defM}
{\cal{M}}_{AB} (\lambda) \equiv {{{\cal{D}}_{AB} (\lambda)} \over {D_f
(\lambda)}} .
\end{equation}
Differentiating ${\cal{M}}_{AB}$ twice with respect to $\lambda$ and
using (\ref{evoD}), one finds
\begin{equation}
\label{difM}
{{d^2{\cal{M}}_{AB}} \over {d\lambda^2}} = -{2 \over {D_f}} {{d D_f}
\over {d \lambda}} {{d {\cal{M}}_{AB}} \over {d\lambda}} + \delta
{\cal{T}}_{AC} {\cal{M}}_{CB}.
\end{equation}
With the initial conditions {\boldmath${\cal{M}}$}$(\lambda) |_{\lambda=0} =
${\boldmath${\cal{I}}$} and $d${\boldmath${\cal{M}}$}
$(\lambda)/{d\lambda} |_{\lambda=0} = ${\boldmath${\cal{O}}$}, \cite{se94} 
the last equation can be written in the integral form 
\begin{equation}
\label{Mab}
{\cal{M}}_{AB} (\lambda) = \delta_{AB} + \int_0^{\lambda} d \lambda'
{{D_f(\lambda-\lambda') D_f(\lambda')} \over {D_f(\lambda)}} \delta
{\cal{T}}_{AC}(\lambda') {\cal{M}}_{CB} (\lambda'). 
\end{equation}
This is the general form of the evolution equation of the lensing magnification
matrix relative to the Friedmann distance in the multiple gravitational
lensing theory.~\cite{sc92}

Next, we derive an evolution equation of the lensing magnification
matrix relative to the Dyer-Roeder distance in the same manner as
above. 
First, we rederive the Dyer-Roeder distance from (\ref{evoD}) 
under the following assumptions:~\cite{dr72,dr73}
\begin{enumerate}
\renewcommand{\labelenumi}{(\Roman{enumi})}
\item The intergalactic space where the light rays propagate has a
uniform matter density $\tilde{\alpha} \rho_b$, where $0 \leq
\tilde{\alpha} \leq 1$.
\item The shear of the bundle of light rays can be ignored.
\item The relation between the affine parameter and the redshift is still
given by that in the homogeneous background universe.
\end{enumerate}
From assumptions (I) and (II), the optically tidal matrix
{\boldmath${\cal{T}}$} becomes
\begin{equation}
\label{taualp}
\mbox{{\boldmath${\cal{T}}$}} = 
\mbox{{\boldmath${\cal{T}}$}}^{\tilde{\alpha} (0)} =
\left[
-K + \left( 1-\tilde{\alpha} \right) {3\over 2} {{\Omega_0} \over a} 
\right]
\mbox{{\boldmath${\cal{I}}$}},
\end{equation}
and the Jacobi differential equation reduces to the scalar form
\begin{equation}
\label{DRdif}
{{d^2 \cal{D}} \over {d \lambda^2}} = {\cal{T}}^{\tilde{\alpha} (0)}
\cal{D}.
\end{equation}
It was shown in Ref.~\citen{sa93} that, by using the assumption (III),
the last equation can be shown to be equivalent to the usual
Dyer-Roeder differential equation. 
Therefore the solution of (\ref{DRdif}) is, of course, the Dyer-Roeder 
distance, and we denote it as $D_{\tilde{\alpha}}$.

We also assume that the matter density in the universe can be
decomposed into a uniform part and a clumpy part as
\begin{eqnarray}
\label{decomp}
\rho 
&=& \rho_{un} + \rho_{cl} \nonumber \\
&=& \tilde{\alpha} \rho_b + \rho_{cl},
\end{eqnarray}
with $\langle \rho_{cl} \rangle = (1-\tilde{\alpha}) \rho_b$. 
By using the above definitions, the optically tidal matrix can be
rewritten as 
\begin{equation}
\label{taual}
\mbox{{\boldmath${\cal{T}}$}} = 
{\cal{T}}^{\tilde{\alpha} (0)} 
\mbox{{\boldmath${\cal{I}}$}} +
\mbox{{\boldmath${\delta \cal{T}}$}}^{cl},
\end{equation}
where {\boldmath${\delta \cal{T}}$}$^{cl}$ has the same form as the second
term of the right-hand side of (\ref{tid}), but the Newtonian
potential $\phi$ is replace by $\phi^{cl}$, which is determined by the
following Poisson equation:
\begin{equation}
\label{poicl}
 \Delta^{(3)} \phi^{cl}= 4 \pi G \rho_{cl} a^2.
\end{equation}
It is again natural to define the lensing magnification matrix
relative to the Dyer-Roeder distance as
{\boldmath${\cal{M}}$}$^{\tilde{\alpha}}
\equiv${\boldmath${\cal{D}}$}$/D_{\tilde{\alpha}}$. Performing the
same procedure as in the case of (\ref{defM}) and below, we obtain
\begin{equation}
\label{Mabal}
{\cal{M}}^{\tilde{\alpha}}_{AB} (\lambda) = \delta_{AB} +
\int_0^{\lambda} d \lambda'
{{D_{\tilde{\alpha}}(\lambda-\lambda') D_{\tilde{\alpha}}(\lambda')}
\over {D_{\tilde{\alpha}}(\lambda)}} \delta
{\cal{T}}^{cl}_{AC}(\lambda') {\cal{M}}^{\tilde{\alpha}}_{CB}
(\lambda'). 
\end{equation}
The last integral equation of the lensing magnification matrix relative to
the Dyer-Roeder distance has the same form as that of one relative to
the Friedmann distance, (\ref{Mab}), but the distances and
gravitational potential are replaced by $D_{\tilde{\alpha}}$ and
$\phi^{cl}$. 

As a consequence of the definitions of the lensing magnification matrices
relative to the background Friedmann and Dyer-Roeder distances, the
relation between these two is obviously
\begin{equation}
\label{relat}
\mbox{{\boldmath${\cal{M}}$}} = {{D_{\tilde{\alpha}}} \over {D_f}}
{\mbox{{\boldmath${\cal{M}}$}}}^{\tilde{\alpha}}.  
\end{equation}
The image magnification of a point-like source is given by the
inverse of the determinant of the magnification matrix. We denote the
magnification relative to the standard Friedmann distance as $\mu_f$
and to the Dyer-Roeder distance as $\mu_{\tilde{\alpha}}$; i.e.,
$\mu_f=|\det${\boldmath${\cal{M}}$}$|^{-1}$ and $\mu_{\tilde{\alpha}}
= |\det${\boldmath${\cal{M}}$}$^{\tilde{\alpha}}|^{-1}$,
respectively. From the definitions and the relation (\ref{relat}),
they are related by 
\begin{equation}
\label{relmu}
\mu_f = \left( {{D_f} \over {D_{\tilde{\alpha}}}} \right)^2
\mu_{\tilde{\alpha}}.
\end{equation}
It is important to note that the relation (\ref{relmu}) itself is well 
known, and has been used in gravitational lensing theory.
However, the relation was, conventionally, derived by the averaging of
magnifications over an ensemble of sources based on an argument of
flux conservation.~\cite{sc92} Strictly speaking, the conventional
relation is written as \cite{sc92,eh86} 
\begin{equation}
\label{conv}
\langle \mu_f \rangle = \left( {{D_f} \over {D_{\tilde{\alpha}}}} \right)^2
\langle \mu_{\tilde{\alpha}} \rangle,
\end{equation}
where $\langle \rangle$ represents an ensemble average over sources at
the same redshift. 
It is, therefore, not clear whether the relation is true for each
individual ray bundle. It should be emphasized that, as we have shown
above, the relation comes from the natural consequence of the
definitions of the lensing magnification matrices. Accordingly, the
relation is true for each ray bundle. 
 
\section{Summary}
In the present paper, we have shown that the integral equations of
the lensing magnification matrix are obtained from the null geodesic
deviation equation with the natural definitions for the magnification
matrices. The integral equations (\ref{Mab}) and (\ref{Mabal}) may be
regarded as the general form for the evolution equation of the lensing
magnification matrix in the multiple gravitational thin lensing
theory.~\cite{sc92} 
Therefore our definitions of the lensing magnification
matrices evidently give the general form for that in the gravitational
thin lensing theory. As a natural consequence of the definitions of
the lensing magnification matrices, the relation between the
magnification relative to the Friedmann distance and that relative to
the Dyer-Roeder distance (\ref{relmu}) is obtained. It should be noted
that the averaging of magnifications over an ensemble of sources was
not performed to derive the relation (\ref{relmu}). Accordingly, we
found that the relation is true for each individual ray bundle as well
as the ensemble average of magnifications of sources at the same
redshift. 

\section*{Acknowledgements}
The author would like to thank Professor T.~Futamase for valuable
discussions and carefully reading the manuscript. He would also like to
thank Professor P.~Schneider, Dr.~M.~Hattori, Dr.~P.~Premadi and
M.~Takada for fruitful discussions. 


\end{document}